%% file: main-icc.tex
\documentclass[conference]{IEEEtran}
\IEEEoverridecommandlockouts

\input{src/preamble}

\newcommand\copyrightnotice{%
\begin{tikzpicture}[remember picture,overlay]
\node[anchor=south,yshift=20pt] at (current page.south)
  {\parbox{0.8\textwidth}{\centering\footnotesize
  This work has been submitted to the IEEE for possible publication. Copyright may be transferred without notice, after which this version may no longer be accessible.}};
\end{tikzpicture}}

\usepackage[top=0.7in,left=0.625in,right=0.625in,bottom=1.04in]{geometry}

\begin{document}
\bstctlcite{IEEEabrv:opt}

\title{Flux: Optimal Scheduling of Optical Circuit Switches for LLM Training}

\author{
    \IEEEauthorblockN {
        Arno Troch\IEEEauthorrefmark{1}, %\orcidlink{0009-0005-0050-7922},
        Seyyidahmed Lahmer\IEEEauthorrefmark{2}, %\orcidlink{},
        Abubakr Nada\IEEEauthorrefmark{2}, %\orcidlink{},
        Roger Sanchez-Vital\IEEEauthorrefmark{1}, %\orcidlink{0000-0003-0976-5766}
        Jeroen Famaey\IEEEauthorrefmark{1}, %\orcidlink{0000-0002-3587-1354}
        Michael Peeters\IEEEauthorrefmark{2} %\orcidlink{0000-0002-1499-3782},
    }

    \IEEEauthorblockA {
        \IEEEauthorrefmark{1}\textit{IDLab, University of Antwerp - imec, Antwerp, Belgium}. \{firstname.lastname\}@uantwerpen.be\\
        \IEEEauthorrefmark{2}\textit{imec, Leuven, Belgium}. \{firstname.lastname\}@imec.be\\
    }
    \thanks{This research was funded in part by the European Union’s Horizon Europe research and innovation programme under the Marie Skłodowska-Curie grant agreement No. 101282240.}
}

\maketitle

\begin{abstract}
\acrfull{OCS} offers high bandwidth density and energy efficiency for LLM training, but incurs a non-negligible reconfiguration delay. Prior work typically schedules optical circuit switches independently of compute, using aggregate traffic demand to determine which circuits to provision and when. We argue that this separation creates a fundamental inefficiency: reconfigurations that ignore the compute timeline can stall communication, resulting in low circuit utilization and large buffer requirements. In this paper, we present \sys, a scheduler that optimally manages optical circuit switches based on the structure of the entire workload. \sys remains effective across a wide range of switching speeds by reusing circuits and amortizing reconfiguration delay behind compute and communication. We show that \sys can reduce training iteration time by more than $10\times$ and peak \acrshort{NIC} buffer requirements by more than three orders of magnitude compared to traditional periodic schedulers. 
\end{abstract}

\copyrightnotice

%=========================================================
\section{Introduction}
\label{sec:intro}
%=========================================================

\gls{LLM} training runs at massive scale. Modern training systems rely on hundreds of thousands of accelerators with frequent bandwidth-intensive collective communication~\cite{gangidi_2024_RDMAEthernetDistributed,qian_2024_AlibabaHPN}. As models grow, these communication operations have become a primary bottleneck. Traditional electrical packet-switched networks are approaching fundamental limits in bandwidth density and power efficiency~\cite{gholami_2024_AIMemoryWall}. \gls{OCS} offers deterministic, high-bandwidth connections without queuing delays or congestion~\cite{poutievski_2022_JupiterEvolving,addanki_2023_Mars}. Unlike packet switching, where individual packets are multiplexed across electrical switches, an optical circuit switch establishes direct optical paths between endpoints, without per-packet buffering or arbitration. However, \gls{OCS} introduces a non-negligible reconfiguration overhead when changing the network topology, so reconfigurations need to be intelligently scheduled.

To manage this overhead, prior work typically treats circuit scheduling as a purely networking problem. Approaches ranging from one-shot static topologies to periodic reconfiguration derive their schedules from an aggregate traffic matrix~\cite{khani_2021_SiPML,liu_2015_SchedulingTechniquesHybrid,wu_2025_ACTINA,wang_2023_TopoOpt}. We argue that this assumption is a poor fit for modern \gls{LLM} training workloads. A traffic matrix sums communication volume over a time window. Therefore, it discards the strict execution order of the training loop. \gls{LLM} training, by contrast, is highly structured and deterministic, since communication operations occur at precise moments and depend on the completion of preceding compute and communication steps~\cite{li_2024_UnderstandingCommunicationCharacteristics}.

Furthermore, separating communication scheduling from compute creates an inefficiency. Schedulers that rely on traffic matrices cannot capture compute-communication overlap, which means that the network may be performing large transfers while accelerators sit idle waiting for data. Producing data without first confirming that the required circuit is available also increases end-host buffer requirements. This means that minimizing network completion time in isolation does not necessarily minimize the end-to-end iteration time of the training workload.

Unlike existing \emph{demand-aware} approaches, which typically generate a schedule from the aggregate traffic volume a specific workload produces, we argue that circuit scheduling must directly account for the temporal structure of the entire workload. By taking into account the temporal dependencies of the workload, we can more effectively reuse circuits and hide reconfiguration latency behind compute phases.

\begin{figure*}[!t]
    \centering
    \includegraphics[width=0.98\textwidth]{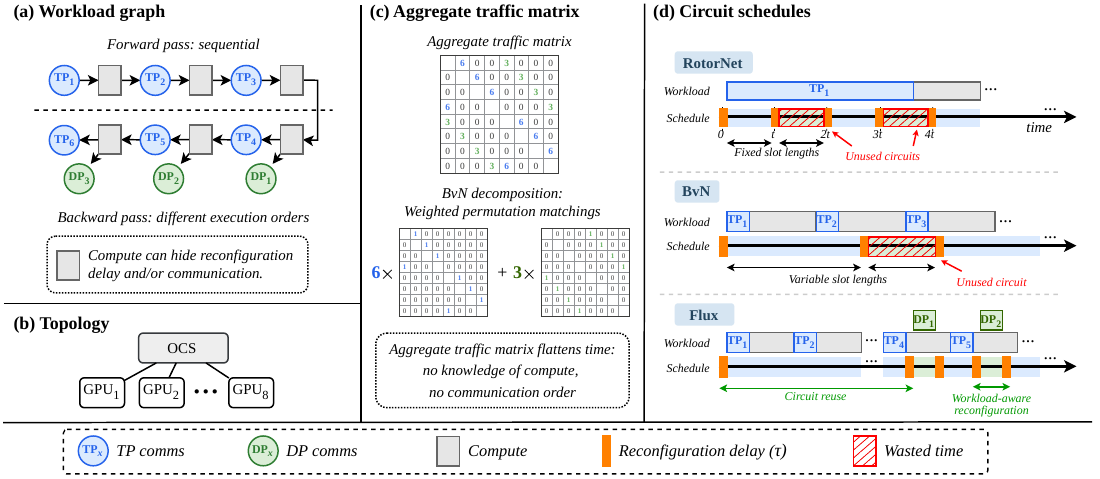}
    \caption{Depiction of how \sys differs from existing circuit scheduling approaches. (a) an example training step (TP + DP) as a graph of compute and communication steps with dependencies (b) the network topology with 1 \acrshort{OCS} switch and 8 \acrshortpl{GPU}; (c) an aggregate traffic matrix of the workload that records total communication volume but discards the graph ordering; (d) three circuit schedules for this workload. RotorNet and BvN-based schedules reconfigure on a fixed period regardless of the workload's state, stalling when the active circuit is wrong. \sys instead derives its schedule from the graph, reusing circuits and amortizing reconfiguration behind compute instead.}
    \label{fig:dag_schedule}
\end{figure*}

To demonstrate this, we introduce \sys, a formalization of workload-aware circuit scheduling as a \gls{MILP} that consolidates compute and communication into a single circuit schedule. We show that \sys reduces training iteration time and buffer requirements by producing schedules that naturally lead to circuit reuse, compute-communication overlap, and amortization of the reconfiguration delay. We make the following contributions:

\begin{itemize}[leftmargin=1.2em]
    \item We show why aggregate traffic matrix scheduling is not an optimal fit for the deterministic, temporally structured nature of LLM training, and motivate the need for \textit{workload-aware circuit scheduling}.
    \item We design \sys, a formalization of workload-aware circuit scheduling as a \gls{MILP} that takes into account reconfiguration delay and workload dependencies.
    \item We evaluate \sys against traditional periodic schedulers and find that it consistently reduces iteration time across a range of reconfiguration delays, while also significantly reducing the \gls{NIC} buffer requirements.
\end{itemize}
\noindent

The remainder of the paper is organized as follows. Section~\ref{sec:related-work} contrasts with related work, Section~\ref{sec:motivation} motivates workload-aware scheduling, Section~\ref{sec:milp} presents the \gls{MILP} formulation, Section~\ref{sec:evaluation} reports the evaluation, and Section~\ref{sec:conclusion} concludes.

%=========================================================
\section{Related Work}
\label{sec:related-work}
%=========================================================

Circuit switching strategies have been widely studied, whether for general datacenter use~\cite{ballani_2020_Sirius,mellette_2017_RotorNet,liu_2015_SchedulingTechniquesHybrid,addanki_2023_Mars} or for distributed training workloads~\cite{jouppi_2023_TPUV4,wu_2025_ACTINA,poutievski_2022_JupiterEvolving,khani_2021_SiPML,wang_2023_TopoOpt}. However, many of these approaches do not exploit the deterministic and temporal nature of \gls{LLM} training, the interaction between compute and communication, and the cost of reconfiguration.

\textbf{Workload-aware circuit scheduling.} Existing circuit scheduling approaches typically rely on deriving a schedule from an aggregate traffic demand matrix, obtained by summing or averaging communication volume over some time window (e.g., a training iteration)~\cite{khani_2021_SiPML,liu_2015_SchedulingTechniquesHybrid,wu_2025_ACTINA,wang_2023_TopoOpt}. For \gls{LLM} training, where compute and communication operations have a specific execution order, aggregating demand over time discards this ordering. Additionally, most of this work treats scheduling as a purely networking problem, decoupled from compute, missing opportunities to overlap communication with compute. In contrast, \sys schedules circuit reconfigurations and communication while explicitly taking into account when compute occurs.

\textbf{Complexity of reconfiguration.} Circuit reconfiguration induces a non-negligible reconfiguration delay, and its cost varies substantially across switching technologies, from nanoseconds~\cite{ballani_2020_Sirius} to milliseconds~\cite{urata_2022_MissionApollo}. A large fraction of existing work does not consider this delay as an explicit parameter of the scheduling problem, instead assuming it to be negligible or fixing it to a value tied to a particular hardware platform. Treating reconfiguration delay as a fixed constant rather than a first-class parameter of the problem obscures the trade-off between the benefit of adapting the topology and the overhead incurred by doing so. Since \sys takes the reconfiguration delay as a parameter and is aware of the workload structure, it naturally creates opportunities for circuit reuse and amortizing the reconfiguration cost.

%=========================================================
\section{Beyond Aggregate Traffic Matrices}
\label{sec:motivation}
%=========================================================

\gls{OCS} introduces a non-negligible reconfiguration delay, which makes deciding which topologies to transition to and when non-trivial. A natural approach to managing this overhead has been to offload scheduling to controllers that monitor demand, or to rely on periodic reconfigurations. In most cases, these systems build their topologies around an aggregate traffic matrix: a matrix where entry $(i,j)$ gives the total traffic volume sent from source $i$ to destination $j$ over a time window. This creates a fundamental limitation: a traffic matrix flattens time. It views the workload as a bulk data transfer problem rather than a sequence of pipelined transfers. This means that while the total communication volume is satisfied, the temporal structure of the \gls{LLM} training iteration is completely lost.

Prior work often attempts to alleviate this gap through intelligent matrix decomposition. For example, a common strategy is to isolate large flows for dedicated optical circuits, while allowing smaller latency-sensitive flows to bypass the reconfiguration penalty via multi-hop routing or even a separate packet-switched network~\cite{liu_2015_SchedulingTechniquesHybrid}. Alternative approaches heavily rely on \gls{BvN} decomposition to partition the demand matrix into a convex combination of permutation matrices~\cite{khani_2021_SiPML,liu_2015_SchedulingTechniquesHybrid,wu_2025_ACTINA,wang_2023_TopoOpt}. However, these techniques still operate on a flattened, aggregate view of demand: partitioning or decomposing the traffic matrix changes how demand is expressed. None of them recover the execution order of the underlying workload, which is especially important for \gls{LLM} training.

We illustrate this disconnect in Fig.~\ref{fig:dag_schedule}. Consider a training step distributed across eight \glspl{GPU} using a mix of \gls{DP}~\cite{rajbhandari_2020_ZeRO} and \gls{TP}~\cite{shoeybi_2020_MegatronLM}. The workload can be represented as a \gls{DAG}, where the vertices represent compute and communication operations, and edges indicate dependencies. As illustrated in Fig.~\ref{fig:dag_schedule}\hyperref[fig:dag_schedule]{c}, the corresponding traffic matrix simply records the total traffic volume exchanged, but not the order in which traffic becomes available. In the example workload graph (cf. Fig.~\ref{fig:dag_schedule}\hyperref[fig:dag_schedule]{a}), the \gls{DP} communication operations become available only in the backward pass, while several \gls{TP} exchanges and computations have to be executed beforehand. Therefore, scheduling these communication operations out of order would stall the entire workload. This highlights a critical flaw: optimizing for aggregate demand can actively harm the application's progress.

Furthermore, this time-agnostic view also eliminates opportunities for compute-communication overlap. Modern distributed training heavily relies on this overlap to amortize network costs. As shown in the example schedules in Fig.~\ref{fig:dag_schedule}\hyperref[fig:dag_schedule]{d}, a workload-aware scheduler such as \sys is capable of actively aligning circuit reconfigurations and communication with ongoing compute phases. If we isolate the network schedule from the compute dependencies, we lose the ability to effectively pipeline these operations.

The same figure further illustrates how RotorNet~\cite{mellette_2017_RotorNet} and a \gls{BvN}-based schedule~\cite{liu_2015_SchedulingTechniquesHybrid} lead to substantial inefficiencies. RotorNet is \emph{demand-oblivious} because it rotates through a fixed set of circuits with a fixed reconfiguration period. The \gls{BvN}-based schedule is \emph{demand-aware} since it rotates through a fixed set of circuits (one per permutation matrix), but with varying reconfiguration periods based on the weight of each permutation matrix. These schedules trigger unnecessary circuit reconfigurations, since they cannot anticipate the workload's state. \sys, by contrast, reduces the total reconfiguration cost by reusing circuits and amortizing the reconfiguration delay behind compute.

This exposes a major difference in perspective. Rather than simply satisfying the communication demand, we should aim to schedule that same communication demand to finish exactly when the compute needs it. This change in perspective is the main motivation for this work, and suggests a need to revisit how to schedule circuits for \gls{LLM} training.

%=========================================================
\section{\sys: Circuit Scheduling as an Optimization Problem}
\label{sec:milp}
%=========================================================

\begin{table}[t]
    \centering
    \caption{Notation used throughout the problem formulation.}
    \begin{tabular}{ll}
    \toprule
    \textbf{Symbol} & \textbf{Definition} \\
    \midrule
         $G = (V, E)$ & Workload graph \\
         $N=\{1,\dots,n\}$ & Accelerators \\
         $S=\{1,\dots,s\}$ & Switches \\
         $(s_v, d_v)$ & src, dst pair of $v \in V$ \\
         $\mathcal{C}_v$ & Direct children of $v \in V$ \\
         $\mathcal{D}_v$ & Descendants of $v \in V$ \\
         $\mathcal{A}_v$ & Ancestors of $v \in V$ \\
         $\mathcal{P}_v$ & Direct parents of $v \in V$ \\
         $p_v$ & Transmission time of $v \in V$ \\
         $c_{u,v}$ & Compute time between $u$ and $v$ \\
         $\tau$ & Reconfiguration delay \\
     \bottomrule
    \end{tabular}
    \label{tab:placeholder}
\end{table}

In this section, we present \sys, a formalization of workload-aware circuit scheduling as an \gls{MILP}, which captures the mix of discrete circuit reconfiguration decisions and continuous timing constraints. The formulation takes two inputs: a system of $n$ accelerators interconnected through an \gls{OCS} network and a workload (a specific \gls{LLM} training iteration).

\paragraph{System model} We consider a set $N=\{1,\dots,n\}$ of $n$ accelerators (e.g., \glspl{GPU}) interconnected through a set $S=\{1,\dots,s\}$ of $s$ identical optical circuit switches. Each accelerator has exactly one physical link to each of the $s$ switches, so it can maintain up to $s$ simultaneous bidirectional circuits to other accelerators. Each switch is modeled as an optical $n \times n$ crossbar: at any point in time, it can realize a set of simultaneous point-to-point circuits, with the only restriction being that no two circuits may share an endpoint at that switch. Setting up or changing a circuit implies a reconfiguration delay $\tau$ \si{\second}.

\paragraph{Workload model} As mentioned in Section~\ref{sec:motivation}, a distributed \gls{LLM} training iteration has a highly temporal structure consisting of compute and communication operations and their dependencies. Let $G$ be a \gls{DAG} $G=(V,E)$ where each vertex $v \in V$ is an atomic point-to-point send between two accelerators and each edge $(u,v) \in E$ encodes a dependency: $u$ must fully complete before $v$ can begin. To encode the compute operations from the workload in the graph, each edge $(u,v) \in E$ additionally carries a non-negative weight $c_{u,v} \geq 0$, which represents the amount of compute time that must elapse between two send tasks $u,v \in V$. Each send task $v \in V$ has a fixed source-destination pair $(s_v, d_v) \in N \times N$, and a transmission time $p_v$ \si{\second} that captures how long the corresponding send occupies the link (e.g., message size divided by link capacity).

%%%%%%%%%%%%%%%%%%%%%%%%%%%%%%
\subsection{Objective and Constraints} 
Given a workload graph and a system, we make two decisions per send task: at what time it starts, and on which switch it is executed. We introduce a continuous decision variable $t_v$ for the start time of each task $v \in V$. 

\paragraph{Objective function} 
The objective function minimizes the total completion time of the workload graph:
\begin{equation}
    \min \; \max_{v \in V} \left( t_v + p_v \right).
\label{eq:objective-minmax}
\end{equation}

\paragraph{Switch assignment} 
To capture on which switch a certain send task $v$ will be executed, we introduce a binary decision variable $x_{v,s} \in \{0,1\}$, which equals $1$ if and only if $v$ is assigned to switch $s \in S$. Evidently, every task must be assigned to exactly one switch:
\begin{equation}
    \sum_{s \in S} x_{v,s} = 1 \qquad \forall v \in V.
\label{eq:assignment}
\end{equation}

\paragraph{DAG dependencies}
The structure of the workload graph directly constrains the start times. If task $u$ is a direct parent of task $v$, then $v$ cannot start before $u$ has finished transmitting and any intermediate compute has completed:
\begin{equation}
    t_v \geq t_u + p_u + c_{u,v} \qquad \forall (u,v) \in E.
\label{eq:precedence}
\end{equation}
These constraints fix the execution order of every pair of tasks with a direct edge in the graph, independently of which switch they are assigned to.

The remaining constraints capture the interactions between send tasks $u,v$ that are sent on the same switch. More specifically, these constraints capture the physical limitations of circuit switching related to reconfiguration. When two tasks $u,v$ share an endpoint (i.e., they share a source or destination) and are scheduled on the same switch, there should be a reconfiguration delay $\tau$ between them. Before formally defining this behavior in constraints, we first define a helper set of \emph{potentially conflicting pairs} that collects all pairs of distinct tasks with at least one endpoint in common:
\begin{displaymath}
\begin{aligned}
    \mathcal{K} = \bigl\{ (u,v) \;\big|\;\; & u, v \in V,\; u \neq v,\; \\
    & s_u{=}s_v \,\vee\, d_u{=}d_v \,\vee\, s_u{=}d_v \,\vee\, d_u{=}s_v \bigr\}.
\end{aligned}
\end{displaymath}
For each pair of tasks, a reconfiguration may be necessary if the tasks were to be scheduled on the same switch. We define the reconfiguration penalty incurred when task $v$ executes after task $u$ on the same switch:
\begin{displaymath}
    R_{u,v} =
    \begin{cases}
        0, & \text{if } (s_u, d_u) = (s_v, d_v), \\
        \tau, & \text{otherwise.}
    \end{cases}
\end{displaymath}
If the later task reuses the exact same circuit, there is no need for circuit reconfiguration. Otherwise, the switch must be reconfigured and the start time of the later task is delayed by $\tau$.
 
Whether the relative execution order (i.e., $u$ before $v$ or $v$ before $u$) of a conflicting pair $(u, v)$ is already known depends on the \gls{DAG}. We write $u \prec_G v$ (\emph{precedence}) if and only if $v$ is a descendant of $u$ in $G$, in which case $u$ always executes before $v$. Similarly, we write $u \parallel_G v$ (\emph{incomparability}) if and only if neither task is an ancestor or descendant of the other, in which case their relative execution order becomes a scheduling decision. Based on this distinction, we partition the conflicting pairs into two disjoint sets:
\begin{displaymath}
    \begin{aligned}
        \mathcal{K}^{\prec} &= \bigl\{ (u,v) \in \mathcal{K} \;\big|\; u \prec_G v \bigr\}, \\
        \mathcal{K}^{\parallel} &= \bigl\{ \{u,v\} \;\big|\; (u,v) \in \mathcal{K},\; u \parallel_G v \bigr\}.
    \end{aligned}
\end{displaymath}
 
\paragraph{Fixed-order reconfiguration}
For pairs in $\mathcal{K}^{\prec}$, the order is known, but the precedence constraints from Equation~\eqref{eq:precedence} do not yet account for the reconfiguration delay that arises when both tasks are placed on the same switch without reusing the same circuit. For every such pair and every switch, we add:
\begin{equation}
    \begin{aligned}
        \forall& (u,v) \in \mathcal{K}^{\prec},\; \forall s \in S:\\
        &t_v \geq t_u + p_u + R_{u,v} - M \left( 2 - x_{u,s} - x_{v,s} \right).
    \end{aligned}
    \label{eq:fixed-order}
\end{equation}
In this set of constraints, we use the Big-M method~\cite{bazaraa_1977_LinearProgrammingNetwork}, where we choose a sufficiently large constant $M$ to deactivate conditional constraints, allowing us to write the constraints linearly. When $u$ and $v$ are placed on different switches, at least one of $x_{u,s}, x_{v,s}$ equals $0$ and the subtracted Big-M term deactivates the constraint. When both tasks share switch $s$, the constraint enforces the reconfiguration overhead $R_{u,v}$. Practically speaking, we set the value of $M$ to be the equivalent of scheduling all compute and communication tasks in sequence and reconfiguring between each task:
\begin{displaymath}
    M = |V| \cdot \tau + \sum_{v \in V} p_v + \sum_{(u, v) \in E} c_{u,v}.
\end{displaymath}

\paragraph{Free-order reconfiguration}
For incomparable pairs in $\mathcal{K}^{\parallel}$, the scheduler must also decide their relative execution order. If $u$ and $v$ are placed on the same switch, then either $u$ finishes (plus any required reconfiguration) before $v$ starts, or vice versa. To express this linearly, we introduce, for each unordered pair $\{u,v\} \in \mathcal{K}^{\parallel}$, a binary sequencing variable $y_{u,v} \in \{0,1\}$ that equals $1$ if task $u$ is scheduled before task $v$, and $0$ otherwise. For every pair and every switch, we impose:
\begin{equation}
    \begin{aligned}
        \forall& \{u,v\} \in \mathcal{K}^{\parallel},\; \forall s \in S: \\
        &t_v \geq t_u + p_u + R_{u,v} - M \left( 3 - x_{u,s} - x_{v,s} - y_{u,v} \right) \\
        &t_u \geq t_v + p_v + R_{v,u} - M \left( 2 - x_{u,s} - x_{v,s} + y_{u,v} \right).
    \end{aligned}
    \label{eq:free-order}
\end{equation}
Whenever $u$ and $v$ are not both assigned to switch $s$, both inequalities are deactivated since tasks on different switches never interfere. When $x_{u,s} = x_{v,s} = 1$, then $y_{u,v}$ controls which of the two constraints is activated.

%=========================================================
\section{Evaluation}
\label{sec:evaluation}
%=========================================================
We empirically evaluate whether \emph{workload-aware circuit scheduling} translates into end-to-end improvements under realistic training traces. In particular, we ask three questions: (1) how do different reconfiguration delays impact scheduling? (2) does workload awareness alleviate \gls{NIC} buffering pressure? and (3) does minimizing the number of reconfigurations translate to performance improvements?

To answer these questions, we implement the \gls{MILP} described in Section~\ref{sec:milp} in Gurobi~\cite{gurobi} and use it to compute optimal schedules under our model. We then replay these schedules in a discrete-event network simulator and compare \sys against RotorNet and a \gls{BvN}-based schedule. Our results show that our method consistently achieves the lowest iteration time across the evaluated reconfiguration delays, while significantly reducing peak \gls{NIC} buffer demand by more than three orders of magnitude relative to the other schedulers.

%%%%%%%%%%%%%%%%%%%%%%%%%%%%%%
\subsection{Simulation Setup}
We evaluate \sys using discrete-event network simulations built on an adapted version of ASTRA-sim~\cite{won_2023_ASTRAsim20}. We integrate an internal \gls{OCS} simulator with a modified version of ASTRA-sim to enable circuit switching control. This \gls{OCS} simulator takes as input, for each switch, a sequence of circuit states and the times at which the switch transitions between. For each workload and hardware configuration, we first solve the \gls{MILP} from Section~\ref{sec:milp} with Gurobi to obtain a schedule, then convert it into this input format and simulate it.

\paragraph{Generating workload traces} We generate \gls{LLM} training traces using an in-house workload execution trace generator. This tool generates Chakra~\cite{sridharan_2023_Chakra} traces, a graph-based representation of AI/ML workloads, for a given model and system configuration, which we then feed to ASTRA-sim. For our evaluation, we use traces based on the Llama 3 8B model~\cite{grattafiori_2024_Llama3Herd}. Due to the nature of our approach (i.e., finding a globally optimal solution for the \gls{MILP}), we limit the scale of the evaluation to 16 Transformer layers. Scalability is briefly discussed in Section~\ref{subsec:discussion}.

\paragraph{Hardware and network topology} We model a system setup with 8 \glspl{GPU} and $s=2$ optical circuit switches, following the same network topology as illustrated in Fig.~\ref{fig:dag_schedule}\hyperref[fig:dag_schedule]{b}. Each \gls{GPU} has one \SI{800}{\giga\bps} bidirectional link to each switch, for two links per \gls{GPU} in total. For the compute, we model an Nvidia A100 \gls{GPU} using the hierarchical roofline model from Kelis~\cite{guo_2025_KeepingLargeLanguage}. Each \gls{GPU} is equipped with a \gls{NIC} with two links, one to each \gls{OCS}. We assume that each \gls{NIC} has an unlimited buffer to store generated packets, allowing us to measure the effect of the circuit schedule on peak buffer requirements.

\paragraph{Circuit schedulers} We compare \sys against RotorNet and a \gls{BvN}-based schedule. We implement RotorNet in two variants: a single-hop variant (\verb|RotorNet-Direct|) and a variant using \gls{VLB}. Both variants use a \SI{90}{\percent} duty cycle, meaning that slot length increases with reconfiguration delay. For the \gls{BvN}-based scheduler, we construct a traffic matrix from the workload and decompose it into permutation matrices with corresponding weights. We then use the permutation matrices as circuit configurations and their weights to determine slot lengths. Unlike RotorNet, the slot lengths in the \gls{BvN}-based schedule are independent of the reconfiguration delay.

%%%%%%%%%%%%%%%%%%%%%%%%%%%%%%
\subsection{Performance Results}

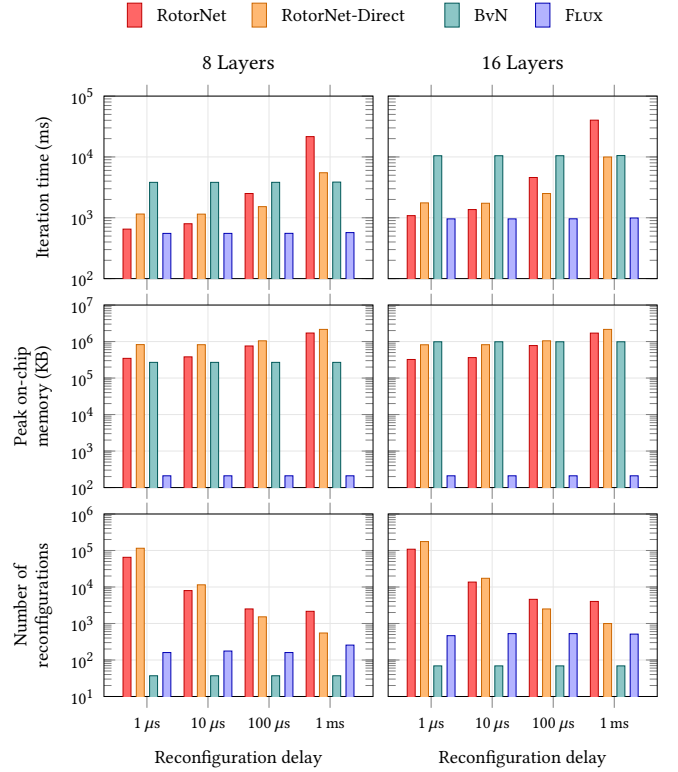
\begin{figure}[t]
\centering
\input{graphics/result-1}
\caption{End-to-end training iteration time (top), peak \acrshort{NIC} buffer usage (middle), and number of reconfigurations (bottom) across different reconfiguration delays for 8 and 16 layers of Llama 3, TP4$\times$DP2.}
\label{fig:performance}
\end{figure}

Fig.~\ref{fig:performance} shows the end-to-end training iteration time, peak buffer requirements, and number of reconfigurations, with reconfiguration delays ranging from \SI{1}{\micro\second} to \SI{1}{\milli\second}. 

The iteration time shows that RotorNet performs well at small reconfiguration delays, where its time slots are smaller and require less waiting for the right circuit. However, the iteration time grows with increasing reconfiguration delay. At higher delays, the demand-aware \gls{BvN} schedule performs better due to its constant slot lengths. Nevertheless, \sys consistently achieves the lowest iteration time, since it schedules circuits while keeping into account the temporal structure of the workload, allowing communication and reconfiguration to be overlapped with compute. These results suggest that our scheduling approach can be effectively used across a wide variety of \gls{OCS} switching technologies, from slower \gls{MEMS}-based switches with millisecond switching times~\cite{urata_2022_MissionApollo} to silicon photonics-based circuit switches with micro- or nanosecond reconfiguration delay~\cite{wang_2025_CoDesignedSiliconPhotonics}.

A major improvement of \sys is in peak \gls{NIC} buffer demand. RotorNet and \gls{BvN} are unaware of the temporal structure of the workload, so communication tasks may become available while there is no suitable circuit to use. These tasks accumulate in the \gls{NIC} packet buffers, resulting in high peak memory usage. In contrast, \sys schedules circuits based on the entire workload structure, keeping buffer requirements low.

Finally, the number of reconfigurations performed in each experiment provides an interesting insight: a system that reconfigures less often does not necessarily perform better. At low reconfiguration delays, RotorNet performs substantially more reconfigurations than \gls{BvN} and \sys, yet achieves competitive iteration times. Similarly, \gls{BvN} consistently performs fewer reconfigurations than our approach, but has longer iteration times and substantially higher memory usage. Therefore, what matters is not how often the network reconfigures, but whether a reconfiguration enables useful circuits.

%%%%%%%%%%%%%%%%%%%%%%%%%%%%%%
\subsection{Discussion}
\label{subsec:discussion}
Our results demonstrate the effectiveness of \emph{workload-aware circuit scheduling} and \sys for \gls{LLM} training. Our approach leverages an exact \gls{MILP} formulation to identify optimal scheduling solutions, allowing us to assess its performance under different training configurations. However, the computational complexity of solving the \gls{MILP} increases with the number of \glspl{GPU} and Transformer layers, making exact optimization challenging for larger training jobs. Therefore, we focus our evaluation on small-scale scenarios, where optimal solutions can be obtained within a reasonable time. Extending the approach to larger deployments is an interesting direction for future work, particularly through the design of heuristics or relaxation techniques that can reduce computational complexity while preserving the performance benefits demonstrated in this study.

Another unresolved problem is how \sys can accommodate \gls{MoE} models. \gls{MoE} models generate non-deterministic communication patterns by routing tokens to different experts~\cite{shazeer_2017_OutrageouslyLargeNeural}. \sys assumes perfect knowledge of communication types and sizes: a realistic assumption for all types of parallelism used in \gls{LLM} training and inference, except \gls{EP} used in \gls{MoE} models.

%=========================================================
\section{Conclusion}
\label{sec:conclusion}
%=========================================================

In this paper, we present \sys, a formalization of \emph{workload-aware circuit scheduling} for \gls{LLM} training workloads. We show that existing traffic matrix-based reconfiguration schemes are not well suited for these temporally structured workloads, where it is important to consider the reconfiguration delay of the system, the temporal dependencies of the workload, and the compute when scheduling the communication. We demonstrate that our approach naturally enables circuit reuse, amortizes reconfiguration and communication costs, and significantly reduces peak \gls{NIC} buffer requirements.

In future work, our objective will be to turn \sys from a powerful formalization into a tool that can be used to schedule problems at any scale. This encompasses the design of intelligent heuristics, support for \gls{MoE} models, and an extensive evaluation campaign.

%\clearpage
\bibliographystyle{IEEEtran}
\bibliography{bib/2026-Flux}

\end{document}

%% file: src/preamble.tex
\usepackage{amsmath}
\usepackage{algpseudocode}
\usepackage[acronym]{glossaries}
\usepackage{booktabs}
\usepackage{hyperref}
\usepackage{enumitem}
\usepackage{microtype}
\usepackage{pdfpages}
\usepackage{siunitx}
\usepackage{subcaption}
\usepackage{tabularx}
\usepackage{textcomp}
\usepackage[]{threeparttable}
\usepackage[colorinlistoftodos,prependcaption,textsize=tiny]{todonotes}
\usepackage{xargs}
\usepackage{xcolor}
\usepackage{pgfplots}
\usepackage{balance}
\usepgfplotslibrary{groupplots}
\usetikzlibrary{calc,patterns}
\pgfplotsset{compat=1.18}

\newcommandx{\inlinetodo}[2][1=]{\todo[inline,caption={\textbf{TODO}},linecolor=red,backgroundcolor=red!25,bordercolor=red,#1]{#2}}
\newcommandx{\margintodo}[2][1=]{\todo[linecolor=blue,backgroundcolor=blue!25,bordercolor=blue,#1]{#2}}
\DeclareSIUnit{\bits}{b}
\DeclareSIUnit{\bps}{bps}
\DeclareSIUnit{\Bps}{Bps}
\DeclareSIUnit{\dBm}{dBm}

\setlist[itemize]{topsep=2pt, itemsep=1pt, parsep=0pt, partopsep=0pt, leftmargin=*}

\newcommand*\myglsentry[1]{%
  \protect\ifglsused{#1}{%
    \glsentryshort{#1}%
  }{%
    \glsentrylong{#1}%
  }%
}
\loadglsentries[acronym]{./src/glossary}
\usepackage{xspace}

\graphicspath{{./graphics/}}

\newcommand{\sys}{\textsc{Flux}\xspace}

\def\BibTeX{{\rm B\kern-.05em{\sc i\kern-.025em b}\kern-.08emT\kern-.1667em\lower.7ex\hbox{E}\kern-.125emX}}

%% file: src/glossary.tex
\newacronym[firstplural=Radio Access Technologies (RATs)]{RAT}{RAT}{Radio Access Technology}
\newacronym{3GPP}{3GPP}{3rd Generation Partnership Program}
\newacronym{5GC}{5GC}{5G Core}
\newacronym{AI}{AI}{Artificial Intelligence}
\newacronym{AIFSN}{AIFSN}{Arbitration Inter-Frame Space Number}
\newacronym{AP}{AP}{Access Point}
\newacronym{BSS}{BSS}{Basic Service Set}
\newacronym{BvN}{BvN}{Birkhoff-von Neumann}
\newacronym{BW}{BW}{Bandwidth}
\newacronym{BWP}{BWP}{Bandwidth Part}
\newacronym{C-ITS}{C-ITS}{Cooperative \myglsentry{ITS}}
\newacronym{C-V2X}{C-V2X}{Cellular \myglsentry{V2X}}
\newacronym{C2CCC}{C2CCC}{CAR 2 CAR Communication Consortium}
\newacronym{CA}{CA}{Carrier Aggregation}
\newacronym{CAM}{CAM}{Cooperative Awareness Message}
\newacronym{CBR}{CBR}{Channel Busy Ratio}
\newacronym{CCAM}{CCAM}{Cooperative, Connected and Automated Mobility}
\newacronym{CEPT}{CEPT}{European Conference of Postal and Telecommunications}
\newacronym{CN}{CN}{Core Network}
\newacronym{CPM}{CPM}{Collective Perception Message}
\newacronym{CPO}{CPO}{Co-Packaged Optics}
\newacronym{DAG}{DAG}{Directed Acyclic Graph}
\newacronym{D2D}{D2D}{die-to-die}
\newacronym{DC}{DC}{Dual-Connectivity}
\newacronym{DCC}{DCC}{Decentralized Congestion Control}
\newacronym{DCM}{DCM}{Dual Carrier Modulation}
\newacronym{DENM}{DENM}{Decentralized Environmental Notification Message}
\newacronym{DL}{DL}{downlink}
\newacronym{DP}{DP}{data parallelism}
\newacronym{DSRC}{DSRC}{Dedicated Short Range Communications}
\newacronym{DWDM}{DWDM}{Dense Wavelength-Division Multiplexing}
\newacronym{E2E}{E2E}{End-to-End}
\newacronym{EDCA}{EDCA}{Enhanced Distributed Channel Access}
\newacronym{eNB}{eNB}{Evolved Node B}
\newacronym{EP}{EP}{expert parallelism}
\newacronym{EPC}{EPC}{Evolved Packet Core}
\newacronym{EPS}{EPS}{Electrical Packet Switching}
\newacronym{ETSI}{ETSI}{European Telecommunications Standards Institute}
\newacronym{EU}{EU}{European Union}
\newacronym{eV2X}{eV2X}{Enhanced \myglsentry{V2X}}
\newacronym{FCC}{FCC}{Federal Communications Commission}
\newacronym{FDM}{FDM}{Frequecy Division Multiplexing}
\newacronym{FFN}{FFN}{Feed Forward Network}
\newacronym{FR1}{FR1}{Frequency Range 1}
\newacronym{FR2}{FR2}{Frequency Range 2}
\newacronym{FSDP}{FSDP}{Fully Sharded \myglsentry{DP}}
\newacronym{FSPL}{FSPL}{Free-Space Path Loss}
\newacronym{FTR}{FTR}{Fluctuating Two-Ray}
\newacronym{FWA}{FWA}{Fixed Wireless Access}
\newacronym{GEMM}{GEMM}{General Matrix Multiplication}
\newacronym{gNB}{gNB}{G Node B}
\newacronym{GPU}{GPU}{Graphics Processing Unit}
\newacronym{GQA}{GQA}{Grouped Query Attention}
\newacronym{HARQ}{HARQ}{Hybrid Automatic Repeat reQuest}
\newacronym{HBM}{HBM}{High Bandwidth Memory}
\newacronym{HD}{HD}{High Definition}
\newacronym{IBSS}{IBSS}{Independent Basic Service Set}
\newacronym{IEEE}{IEEE}{Institute of Electrical and Electronics Engineers}
\newacronym{IoV}{IoV}{Internet of Vehicles}
\newacronym{ISAC}{ISAC}{Integrated Sensing And Communication}
\newacronym{ITS-S}{ITS-S}{ITS-G5 Station}
\newacronym{ITS}{ITS}{Intelligent Transport Systems}
\newacronym{KPI}{KPI}{Key Performance Indicator}
\newacronym{LLM}{LLM}{Large Language Model}
\newacronym{LOS}{LOS}{Line of Sight}
\newacronym{LRR}{LRR}{Long Range Radar}
\newacronym{LTE}{LTE}{Long Term Evolution}
\newacronym{MAC}{MAC}{Medium Access Control}
\newacronym{MC}{MC}{Multi-Connectivity}
\newacronym{MCS}{MCS}{Modulation and Coding Scheme}
\newacronym{MEC}{MEC}{Multi-access Edge Computing}
\newacronym{MEMS}{MEMS}{Micro Electro-Mechanical Systems}
\newacronym{MILP}{MILP}{Mixed Integer Linear Program}
\newacronym{ML}{ML}{Machine Learning}
\newacronym{MLP}{MLP}{Multi-Layer Perceptron}
\newacronym{MME}{MME}{Mobility Management Entity}
\newacronym{MST}{MST}{Minimum Spanning Tree}
\newacronym{mmWave}{mmWave}{millimeter wave}
\newacronym{MoE}{MoE}{mixture-of-experts}
\newacronym{NAS}{NAS}{Non-Access Stratum}
\newacronym{NIC}{NIC}{Network Interface Card}
\newacronym{NLOS}{NLOS}{Non-\myglsentry{LOS}}
\newacronym{NLOSv}{NLOSv}{\myglsentry{NLOS} due to vehicles}
\newacronym{NR-U}{NR-U}{\myglsentry{NR} Unlicensed}
\newacronym{NR}{NR}{New Radio}
\newacronym{NSA}{NSA}{Non-Standalone}
\newacronym{OBU}{OBU}{On-Board Unit}
\newacronym{OCB}{OCB}{Outside the Context of a Basic Service Set}
\newacronym{OCS}{OCS}{Optical Circuit Switching}
\newacronym{OFDM}{OFDM}{Orthogonal Frequency-Division Multiplexing}
\newacronym{OFDMA}{OFDMA}{Orthogonal Frequency-Division Multiple Access}
\newacronym{OSI}{OSI}{Open Systems Interconnection}
\newacronym{P2P}{P2P}{Pedestrian-to-Pedestrian}
\newacronym{PDCP}{PDCP}{Packet Data Convergence Protocol}
\newacronym{PDN}{PDN}{Packet Data Network}
\newacronym{PGW}{PGW}{\myglsentry{PDN} Gateway}
\newacronym{PMN}{PMN}{Perceptive Mobile Network}
\newacronym{PP}{PP}{pipeline parallelism}
\newacronym{ProSe}{ProSe}{Proximity Services}
\newacronym{PSFCH}{PSFCH}{Physical Sidelink Feedback Channel}
\newacronym{QAM}{QAM}{Quadrature Amplitude Modulation}
\newacronym{QoS}{QoS}{Quality of Service}
\newacronym{R2R}{R2R}{\myglsentry{RSU}-to-\myglsentry{RSU}}
\newacronym{RAN}{RAN}{Radio Access Network}
\newacronym{RB}{RB}{Resource Block}
\newacronym{RDCN}{RDCN}{Reconfigurable \myglsentry{DCN}}
\newacronym{RLC}{RLC}{Radio Link Control}
\newacronym{RRC}{RRC}{Radio Resource Control}
\newacronym{RSU}{RSU}{Road Side Unit}
\newacronym{SA}{SA}{Standalone}
\newacronym{SAE}{SAE}{Society of Automotive Engineers}
\newacronym{SB-SPS}{SB-SPS}{Sensing-Based Semi-Persistent Scheduling}
\newacronym{SCH}{SCH}{Secondary Cell Handover}
\newacronym{SCM}{SCM}{Statistical Channel Model}
\newacronym{SCS}{SCS}{Subcarrier Spacing}
\newacronym{SDN}{SDN}{Software-Defined Networking}
\newacronym{SGW}{SGW}{Serving Gateway}
\newacronym{SINR}{SINR}{Signal to Interference plus Noise Ratio}
\newacronym{SL}{SL}{Sidelink}
\newacronym{SNR}{SNR}{Signal to Noise Ratio}
\newacronym{SRR}{SRR}{Short Range Radar}
\newacronym{STA}{STA}{Station}
\newacronym{TDC}{TDC}{Transmit Datarate Control}
\newacronym{TDD}{TDD}{Time Division Duplex}
\newacronym{TDM}{TDM}{Time Division Multiplexing}
\newacronym{TM}{TM}{Traffic Matrix}
\newacronym{ToR}{ToR}{Top-of-Rack}
\newacronym{TP}{TP}{tensor parallelism}
\newacronym{TPC}{TPC}{Transmit Power Control}
\newacronym{TR}{TR}{Technical Report}
\newacronym{TRC}{TRC}{Transmit Rate Control}
\newacronym{TS}{TS}{Technical Specification}
\newacronym{TXOP}{TXOP}{Transmission Opportunity}
\newacronym{UDP}{UDP}{User Datagram Protocol}
\newacronym{UE}{UE}{User Equipment}
\newacronym{UL}{UL}{uplink}
\newacronym{UPA}{UPA}{Uniform Planar Array}
\newacronym{URLLC}{URLLC}{Ultra-Reliable and Low-Latency Communications}
\newacronym{V2D}{V2D}{Vehicle to Device}
\newacronym{V2I}{V2I}{Vehicle-to-Infrastructure}
\newacronym{V2N}{V2N}{Vehicle-to-Network}
\newacronym{V2P}{V2P}{Vehicle-to-Pedestrian}
\newacronym{V2R}{V2R}{Vehicle-to-\myglsentry{RSU}}
\newacronym{V2V}{V2V}{Vehicle-to-Vehicle}
\newacronym{V2X}{V2X}{Vehicle-to-Everything}
\newacronym{VANET}{VANET}{Vehicular Ad Hoc Network}
\newacronym{VLB}{VLB}{Valiant Load Balancing}
\newacronym{WI}{WI}{Work Item}
\newacronym{WLAN}{WLAN}{Wireless Local Area Network}
\newacronym{WLO}{WLO}{Wafer-Level Optics}

%% file: graphics/result-1.tex
\tikzset{
  comp-bvn/.style={
    fill=teal!45,draw=teal!70!black,line width=.35pt
  },
  comp-rotor/.style={
    fill=red!60,draw=red!75!black,line width=.35pt
  },
  comp-direct/.style={
    fill=orange!55,draw=orange!80!black,line width=.35pt
  },
  comp-flux/.style={
    fill=blue!30,draw=blue!70!black,line width=.35pt
  },
  % Overlay the pattern so that the scheduler's fill is preserved.
  layers-sixteen/.style={
    postaction={
      pattern=north east lines,
      pattern color=black!95
    }
  },
}

% Explicit shifts keep each scheduler's two layer counts adjacent.
\newcommand{\compBarWidth}{4.5pt}
\newcommand{\compBarGap}{.5pt}

\pgfplotsset{
  comp-position/.style={
    bar shift={
      (#1)*(\compBarWidth+\compBarGap)
    }
  },
  rotor-eight/.style={
    comp-rotor,comp-position=-3.5
  },
  direct-eight/.style={
    comp-direct,comp-position=-2.5
  },
  bvn-eight/.style={
    comp-bvn,comp-position=-1.5
  },
  flux-eight/.style={
    comp-flux,comp-position=-.5
  },
  rotor-sixteen/.style={
    comp-rotor,layers-sixteen,comp-position=.5
  },
  direct-sixteen/.style={
    comp-direct,layers-sixteen,comp-position=1.5
  },
  bvn-sixteen/.style={
    comp-bvn,layers-sixteen,comp-position=2.5
  },
  flux-sixteen/.style={
    comp-flux,layers-sixteen,comp-position=3.5
  },
}

\begin{tikzpicture}

\begin{groupplot}[
  group style={
    group size=1 by 3,
    vertical sep=.35cm,
  },
  scale only axis,
  width={\dimexpr .800\columnwidth+.20cm\relax},
  height=.285\columnwidth,
  xmin=.6, xmax=4.4,
  xtick={1,2,3,4},
  xticklabels={{$1\,\mu$s},{$10\,\mu$s},
               {$100\,\mu$s},{$1$\,ms}},
  x tick label style={font=\tiny},
  ymode=log,
  ybar,
  /pgf/bar width=\compBarWidth,
  tick label style={font=\tiny},
  label style={font=\scriptsize},
  ylabel style={font=\scriptsize,align=center},
  grid=major,
  major grid style={gray!20,line width=.25pt},
  axis line style={line width=.35pt},
  tick style={line width=.35pt},
  enlarge x limits=.04,
]

% ============================================================
% ROW 1: ITERATION TIME
% ============================================================

\nextgroupplot[
  ylabel={Iteration time (ms)},
  ymin=100, ymax=100000,
  ytick={100,1000,10000,100000},
  yticklabels={$10^2$,$10^3$,$10^4$,$10^5$},
  xticklabels=\empty,
]

\addplot[rotor-eight] coordinates {
  (1,650.828) (2,797.733) (3,2505.124) (4,21571.610)
};
\addplot[rotor-sixteen] coordinates {
  (1,1084.121) (2,1363.971) (3,4592.102) (4,40222.761)
};

\addplot[direct-eight] coordinates {
  (1,1152.295) (2,1147.637) (3,1523.716) (4,5497.657)
};
\addplot[direct-sixteen] coordinates {
  (1,1756.962) (2,1732.137) (3,2503.716) (4,9977.657)
};

\addplot[bvn-eight] coordinates {
  (1,3821.890) (2,3822.232) (3,3825.652) (4,3859.852)
};
\addplot[bvn-sixteen] coordinates {
  (1,10466.528) (2,10467.158) (3,10473.458) (4,10536.458)
};

\addplot[flux-eight] coordinates {
  (1,553.503) (2,553.612) (3,554.592) (4,570.493)
};
\addplot[flux-sixteen] coordinates {
  (1,958.943) (2,958.520) (3,961.580) (4,991.180)
};

% ============================================================
% ROW 2: PEAK ON-CHIP MEMORY
% ============================================================

\nextgroupplot[
  ylabel={Peak buffer\\usage (KB)},
  ymin=100, ymax=10000000,
  ytick={100,1000,10000,100000,1000000,10000000},
  yticklabels={$10^2$,$10^3$,$10^4$,$10^5$,$10^6$,$10^7$},
  xticklabels=\empty,
]

\addplot[rotor-eight] coordinates {
  (1,346449.920) (2,380633.088)
  (3,756233.216) (4,1712110.592)
};
\addplot[rotor-sixteen] coordinates {
  (1,322793.472) (2,363121.664)
  (3,776156.160) (4,1702937.600)
};

\addplot[direct-eight] coordinates {
  (1,825175.040) (2,819826.688)
  (3,1046948.864) (4,2151778.304)
};
\addplot[direct-sixteen] coordinates {
  (1,815708.160) (2,819826.688)
  (3,1046948.864) (4,2151778.304)
};

\addplot[bvn-eight] coordinates {
  (1,268980.224) (2,268980.224)
  (3,268980.224) (4,268980.224)
};
\addplot[bvn-sixteen] coordinates {
  (1,985711.616) (2,985711.616)
  (3,985711.616) (4,985711.616)
};

\addplot[flux-eight] coordinates {
  (1,209.920) (2,209.920) (3,209.920) (4,209.920)
};
\addplot[flux-sixteen] coordinates {
  (1,209.920) (2,209.920) (3,209.920) (4,209.920)
};

% ============================================================
% ROW 3: NUMBER OF RECONFIGURATIONS
% ============================================================

\nextgroupplot[
  ylabel={Number of\\reconfigurations},
  xlabel={Reconfiguration delay},
  ymin=10, ymax=1000000,
  ytick={10,100,1000,10000,100000,1000000},
  yticklabels={$10^1$,$10^2$,$10^3$,$10^4$,$10^5$,$10^6$},
]

\addplot[rotor-eight] coordinates {
  (1,65082) (2,7977) (3,2505) (4,2157)
};
\addplot[rotor-sixteen] coordinates {
  (1,108412) (2,13639) (3,4592) (4,4022)
};

\addplot[direct-eight] coordinates {
  (1,115229) (2,11476) (3,1523) (4,549)
};
\addplot[direct-sixteen] coordinates {
  (1,175696) (2,17321) (3,2503) (4,997)
};

\addplot[bvn-eight] coordinates {
  (1,37) (2,37) (3,37) (4,37)
};
\addplot[bvn-sixteen] coordinates {
  (1,69) (2,69) (3,69) (4,69)
};

\addplot[flux-eight] coordinates {
  (1,160) (2,176) (3,160) (4,256)
};
\addplot[flux-sixteen] coordinates {
  (1,464) (2,528) (3,528) (4,512)
};

\end{groupplot}

% ============================================================
% SHARED SCHEDULER LEGEND
% ============================================================

\coordinate (legendcenter) at (group c1r1.north);

\draw[comp-rotor]
  ([xshift=-3.35cm,yshift=26pt]legendcenter)
  rectangle
  ([xshift=-3.16cm,yshift=35pt]legendcenter);
\node[anchor=west,font=\scriptsize]
  at ([xshift=-3.08cm,yshift=30.5pt]legendcenter)
  {RotorNet};

\draw[comp-direct]
  ([xshift=-1.70cm,yshift=26pt]legendcenter)
  rectangle
  ([xshift=-1.51cm,yshift=35pt]legendcenter);
\node[anchor=west,font=\scriptsize]
  at ([xshift=-1.43cm,yshift=30.5pt]legendcenter)
  {RotorNet-Direct};

\draw[comp-bvn]
  ([xshift=.85cm,yshift=26pt]legendcenter)
  rectangle
  ([xshift=1.04cm,yshift=35pt]legendcenter);
\node[anchor=west,font=\scriptsize]
  at ([xshift=1.12cm,yshift=30.5pt]legendcenter)
  {BvN};

\draw[comp-flux]
  ([xshift=2.05cm,yshift=26pt]legendcenter)
  rectangle
  ([xshift=2.24cm,yshift=35pt]legendcenter);
\node[anchor=west,font=\scriptsize]
  at ([xshift=2.32cm,yshift=30.5pt]legendcenter)
  {\sys};

% Layer legend occupies the former subplot-title space.
\draw[fill=gray!20,draw=black!55,line width=.35pt]
  ([xshift=-1.45cm,yshift=7pt]legendcenter)
  rectangle
  ([xshift=-1.26cm,yshift=16pt]legendcenter);
\node[anchor=west,font=\scriptsize]
  at ([xshift=-1.18cm,yshift=11.5pt]legendcenter)
  {$8$ Layers};

\draw[
  fill=gray!20,draw=black!55,line width=.35pt,
  layers-sixteen
]
  ([xshift=.25cm,yshift=7pt]legendcenter)
  rectangle
  ([xshift=.44cm,yshift=16pt]legendcenter);
\node[anchor=west,font=\scriptsize]
  at ([xshift=.52cm,yshift=11.5pt]legendcenter)
  {$16$ Layers};

\end{tikzpicture}